\documentclass[a4paper,12pt]{article}
\usepackage{cite}
\usepackage{graphicx}
\usepackage[small]{subfigure,epsfig}
\usepackage{indentfirst}
\usepackage {amssymb,amsmath,latexsym,enumerate}
\usepackage{indentfirst}
\usepackage{amsthm}

\usepackage{geometry} 
\numberwithin{equation}{section} \numberwithin{table}{section}
\begin{document}
%\begin{frontmatter}

\title{Elliptic solutions in the H\'{e}non -- Heiles model}

\author{Maria V. Demina and Nikolai A. Kudryashov}

\maketitle

%\address{Department of Applied Mathematics, National Research Nuclear University "MEPhI", 31 Kashirskoe Shosse,
%115409 Moscow, Russian Federation}

\begin{abstract}

Equations of motion corresponding to the H\'{e}non -- Heiles system are considered.
A method enabling one to find all elliptic solutions of an autonomous ordinary differential equation or a system of autonomous ordinary differential equations is described. New families of elliptic solutions of a fourth--order equation related to the H\'{e}non -- Heiles system are obtained. A classification of elliptic solutions up to the sixth order inclusively is presented.

\end{abstract}

%\begin{keyword}
H\'{e}non -- Heiles system, elliptic solutions,
nonlinear ordinary differential equations
%\end{keyword}

%\end{frontmatter}

\section{Introduction}

The H\'{e}non -- Heiles model  belong to one of the most famous and frequently used models in astronomy and some other fields of physics. For example, the H\'{e}non -- Heiles equations arise in description of  a moving star in the axisymmetric potential of the galaxy. The H\'{e}non -- Heiles model is defined by the following Hamiltonian \cite{Henon01}
\begin{equation}
\label{HH_Hamiltonian} H=\frac{1}{2}\left(x_t^2+y_t^2+\alpha_1x^2+\alpha_2y^2\right)+x^2y-\frac{\sigma}{3}y^3.
\end{equation}
Equations of motion corresponding to the Hamiltonian $H$ are of the form
\begin{equation}
\begin{gathered}
\label{HH_equations_of_motion} x_{tt}=-\alpha_1x-2xy,\hfill \\
y_{tt}=-\alpha_2y-x^2+\sigma y^2.\hfill
\end{gathered}
\end{equation}
As is well known,  in the following cases
\begin{equation}
\begin{gathered}
\label{HH_integrable cases}  \sigma=-1,\quad \alpha_2=\alpha_1;\hfill\\
 \sigma=-6,\quad \alpha_2,\alpha_1\quad  \text{arbitrary};\hfill\\
 \sigma=-16,\quad \alpha_2=16\alpha_1,\hfill\\
\end{gathered}
\end{equation}
the system \eqref{HH_equations_of_motion} is integrable \cite{Weiss01, Weiss02, Weiss03, Fordy01, Conte01}. The values of the parameters given in \eqref{HH_integrable cases} can be obtained by means of the Painlev\'{e} methods. It is a remarkable fact that the integrable cases \eqref{HH_integrable cases} of the H\'{e}non -- Heiles model are related to the stationary flows of the Sawada~-- Kotera equation, the fifth--order Korteweg -- de Vries equation, and the Kaup -- Kupershmidt equation \cite{Fordy01}. While scaling similarity solutions of these partial differential equations can be associated with time--dependent generalizations of the integrable H\'{e}non -- Heiles Hamiltonians \cite{Hone02}. B\"{a}cklund transformations for integrable cases are given in articles \cite{Weiss02, Weiss03,Newell01}.   The general solutions of the H\'{e}non -- Heiles equations are found only in integrable cases \cite{Conte01, Cosgrove01, Cosgrove02}, while only several families of exact solutions are known in other cases \cite{Timoshkova01, Vernov01, Vernov02}. A powerful algebro--geometric approach was used to find certain families of elliptic solutions arising in the H\'{e}non -- Heiles model in integrable cases \cite{Eilbeck01, Eilbeck02}. Some simply periodic and elliptic solutions were obtained in articles \cite{Timoshkova01, Vernov01, Vernov02}. A systematic study of all the situations, when one can get exact solution of the H\'{e}non -- Heiles system in explicit form, has not been undertaken yet.

It is a distinguishing property of a wide class of algebraic autonomous nonlinear ordinary differential
equation equations that all their meromorphic solutions are in fact elliptic, rational in $\cot b z$ ($b$ is a constant) or rational in $z$ with $z$ being an independent variable \cite{Eremenko01, Demina01,Demina02}. Explicit expressions and a way one can construct them in closed form were presented in \cite{Demina03, Demina01,Demina02}.   In this article we describe a method, which can be applied to construct all elliptic solutions arising in the H\'{e}non -- Heiles model. We emphasize that this method enables one to find and classify elliptic solutions not only in integrable, but also in non--integrable cases.  We use this approach to make a subsequent investigation of elliptic solutions of a fourth--order equation related to the H\'{e}non -- Heiles system. Let us mention that despite the existence of the general solutions in the integrable cases it is not an easy task to find all reductions of known general solutions to elliptic functions.

This article is organized as follows. In section \ref{Method applied_HH} we give a detailed description of our approach and present a method enabling one to find any elliptic solutions of an autonomous ordinary differential equation. In section  \ref{HH_Loca_Analysis} we consider  the fourth--order equation associated with the H\'{e}non -- Heiles system and study the local behavior of its meromorphic solutions. In sections $4$, $5$, $6$ we classify elliptic solutions of the equation in question restriction ourselves to the cases of elliptic solutions up to the sixth order inclusively.

\section{Method applied} \label{Method applied_HH}

Let us take an algebraic autonomous nonlinear ordinary differential
equation
\begin{equation}
\label{EQN} E[w(t)]=0.
\end{equation}
Our goal is to describe a method, which can be used to obtain all its elliptic solutions in explicit form. An elliptic function is defined as a meromorphic function periodic in two directions. Any elliptic function can be determined by its behavior in a parallelogram of periods. The number of poles in a parallelogram of
periods, counting multiplicity, is called the order of an elliptic function.  If $w(t)$ is such an elliptic solution of equation \eqref{EQN}, then equation in question has the family of elliptic
solutions $w(t-t_0)$ with $t_0$ being an arbitrary constant.
Equation \eqref{EQN} necessary possesses an elliptic solution if it
admits at least one Laurent expansion in a neighborhood of the pole
$t=t_0$.  Without loss of generality we shall build the Laurent series  in
a neighborhood of the point $t=0$
\begin{equation}
\begin{gathered}
\label{LE1}
w(t)=\sum_{k=1}^{p}\frac{c_{-k}}{t^k}+\sum_{k=0}^{\infty}c_kt^k,\quad
0<|t|<\varepsilon.
\end{gathered}
\end{equation}
Here $p$ is the order of the pole $t=0$. The Painlev\'{e} methods allow one to find all Laurent series satisfying equation \eqref{EQN}. The following proposition was given in article \cite{Demina03}.

\textbf{\textit{Proposition.}} \textit{ Suppose Laurent series
\eqref{LE1} with uniquely determined coefficients  satisfies
equation \eqref{EQN}; then this equation admits at most one
meromorphic solution having a pole $t=0$ with Laurent series
\eqref{LE1}.}

Thus we see that equation
\eqref{EQN} may have at most one elliptic solution possessing the
pole $t=0$ with Laurent series \eqref{LE1} provided that all coefficients in \eqref{LE1} are uniquely determined.

Our algorithm for finding elliptic solutions of equation \eqref{EQN}
in closed form is the following. Note that we omit arbitrary
constant $t_0$.

\textit{Step 1.} Perform local singularity analysis around movable
singular points for solutions of equation \eqref{EQN}.

\textit{Step 2.} Select the order $M$ of an elliptic solution $w(t)$ and take $K$
distinct Laurent series
\begin{equation}
\begin{gathered}
\label{LS_K}
w^{(i)}(t)=\sum_{k=1}^{p_i}\frac{c_{-k}^{(i)}}{t^k}+\sum_{k=0}^{\infty}c_k^{(i)}t^k,\quad
0<|t|<\varepsilon_i,\quad i=1, \ldots, K.
\end{gathered}
\end{equation}
from those, found at step 1, in such a way that the following
conditions
\begin{equation}
\begin{gathered}
\label{ElS_Residues} \sum_{i=1}^{K}c_{-1}^{(i)}=0,\quad
\sum_{i=1}^{K}p_i=M.
\end{gathered}
\end{equation}
hold.

\textit{Step 3.} Define the general expression for the elliptic
solution $w(t)$ possessing $K$ poles $a_1$, $\ldots$,  $a_K$ in a
parallelogram of periods in such a way that the Laurent series in a neighborhood of the point $t=a_i$
is $w^{(i)}(t-a_i)$, $i=1$, $\ldots$, $K$ (see \cite{Demina01, Demina02}). In other words, take the following expression
for $w(t)$
\begin{equation}
\begin{gathered}
\label{Ex_Sol_EllipticK}
w(t)=\sum_{i=1}^{K}c_{-1}^{(i)}\zeta(t-a_i)+\left\{\sum_{i=1}^{K}
\sum_{k=2}^{p_i}\frac{(-1)^k
c_{-k}^{(i)}}{(k-1)!}\frac{d^{k-2}}{dt^{k-2}}\right\}\wp(t-a_i)
+\tilde{h}_0.
\end{gathered}
\end{equation}
Here $\wp(t)$ is the Weierstrass elliptic function satisfying the equation
\begin{equation}
\begin{gathered}
\label{Wier} (\wp_t)^2=4\wp^3-g_2\wp-g_3,
\end{gathered}
\end{equation}
$\zeta(t)$ is the Weierstrass $\zeta$--function, $\tilde{h}_0$ is a
constant.

\textit{Step 4.} Find the Laurent series for $w(t)$ given by
\eqref{Ex_Sol_EllipticK} around its poles $a_1$, $\ldots$, $a_K$.
Without loss of generality set $a_1=0$. Introduces notation
$A_i\stackrel{def}{=}\wp(a_i)$, $B_i\stackrel{def}{=}\wp_t(a_i)$,
$i=2$, $\ldots$, $K$, and
\begin{equation}
\begin{gathered}
\label{h0} h_0\stackrel{def}{=}\tilde{h}_0-
\sum_{i=2}^{K}c_{-1}^{(i)} \zeta(a_i)-\sum_{i\in \,I}^{}c_{-2}^{(i)}
\wp(a_i).
\end{gathered}
\end{equation}
Using addition formulae for the functions $\wp$ and
$\zeta$ (see \cite{Demina01, Demina02} and equalities \eqref{Addition_formulae} below) rewrite
expression \eqref{Ex_Sol_EllipticK} as
\begin{equation}
\begin{gathered}
\label{Ex_Sol_EllipticK_AB}
w(t)=\left\{\sum_{i=2}^{K}\sum_{k=2}^{p_i}\frac{(-1)^k
c_{-k}^{(i)}}{(k-1)!}\frac{d^{k-2}}{dt^{k-2}}\right\}\left(\frac14\left[
\frac{\wp_t(t)+B_i}{\wp(t)-A_i}\right]^2-\wp(t)\right)\\
+\sum_{i=2}^{K}\frac{c_{-1}^{(i)}(\wp_t(t)+B_i)}{2\,(\wp(t)-A_i)}+\left\{\sum_{k=2}^{p_{1}}\frac{(-1)^k
c_{-k}^{(1)}}{(k-1)!}\frac{d^{k-2}}{dt^{k-2}}\right\}\wp(t)+ h_0,
\end{gathered}
\end{equation}

\textit{Step 5.} Require the Laurent series found at the fourth step to coincide with the corresponding Laurent series taken at the second stem.  Form a system of algebraic equations.
Add to this system the equations $B_i^2=4A_i^3-g_2A_i-g_3$, $i=2$,
$\ldots$, $K$. The number of equations in the system should be slightly
more than the number of parameters of elliptic solution
\eqref{Ex_Sol_EllipticK_AB} and equation \eqref{EQN}. Solve the
algebraic system for the parameters of the elliptic solution $w(t)$,
i.e. find $h_0$, $g_2$, $g_3$, $A_i$, $B_i$, $i=2$, $\ldots$, $K$.
In addition correlations for the parameters of equation \eqref{EQN}
may arise. If this system is inconsistent, then equation \eqref{EQN}
does not possess elliptic solutions with supposed Laurent expansions
around poles.

\textit{Step 6.} Check--up obtained solutions, substituting them into the
original equation.

Further, we note that equation \eqref{EQN} may possess the Laurent series with arbitrary coefficients. If such series are taken at the second step, then it is convenient to add the arbitrary coefficients to the list of parameters. In addition in the fifth step of the method the algebraic system can be also obtained in the following way: one substitutes all the Laurent series found in the fourth step into the original equation and sets to zero the coefficients at negative and zero powers of $(t-a_i)$ in the resulting equality. This approach uses the fact that an elliptic function without poles is a constant. Suppose expression \eqref{Ex_Sol_EllipticK} is built on the basis of the Laurent series involving those that contain arbitrary coefficients; then one should verify that all the series are in fact distinct, especially if one forms  the algebraic system with the help of the  aforementioned approach.

In order to find the Laurent series at the fourth step of the method we use addition theorems for the functions $\zeta$ and $\wp$ whenever the elliptic solution $w(t)$ possesses more then two distinct  poles inside a parallelogram of periods. In our notation the formulae given by the addition theorems can be written as
\begin{equation}
\begin{gathered}
\label{Addition_formulae}
\zeta(a_i-a_j)=\zeta(a_i)-\zeta(a_j)+\frac{B_i+B_j}{2(A_i-A_j)},\hfill\\
\wp(a_i-a_j)=-A_i-A_j+\frac{(B_i+B_j)^2}{4(A_i-A_j)^2}\hfill\\
\wp_t(a_i-a_j)=-B_i+\frac{(B_i+B_j)(12A_i^2-g_2)}{4(A_i-A_j)^2} -\frac{B_i\left(B_i+B_j\right)^2}{2(A_i-A_j)^3} \hfill
\end{gathered}
\end{equation}
The values $\zeta(a_i)$, $i=2$, $\ldots$, $K$ disappear from the resulting series provided that we use $h_0$ instead of $\tilde{h}_0$ in the series and expression \eqref{Ex_Sol_EllipticK_AB}.

Further, let us obtain the addition formulae for the following situation $\wp(a_i)=\wp(a_j)$, $a_i\neq a_j$. The Weierstrass elliptic function $\wp(t)$ takes each value twice in a parallelogram of periods. Suppose $2\omega_1$, $2\omega_2$ are the basic periods of the Weierstrass elliptic function. We denote by $\omega_3$ the sum $\omega_1+\omega_2$. The double points are $t=0$, $t=\omega_l$, $l=1$, $2$, $3$ with $t=0$ being the double pole. Since the Weierstrass function is even, we see that the following relation is valid $\wp(2\omega_l-t)=\wp(t)$, $l=1$, $2$, $3$. Consequently, in the case $\wp(a_i)=\wp(a_j)$ we obtain $\wp_t(a_i)=-\wp_t(a_j)$ and $\wp(a_i-a_j)=\wp(2a_i)$. Note that the points $a_i$, $a_j$ are simple, this yields $\wp_t(a_i)\neq0$, $\wp_t(a_j)\neq 0$. Applying the L'H\^{o}pital's rule in~\eqref{Addition_formulae}, we get the addition formulae for the case $\wp(a_i)=\wp(a_j)$:
\begin{equation}
\begin{gathered}
\label{Addition_formulae_double}
\zeta(2a_i)=2\zeta(a_i)+\frac{12A_1^2-g_2}{4B_1},\hfill\\
\wp(2a_i)=-2A_i+\frac{(12A_i^2-g_2)^2}{16B_1^2}\hfill\\
\wp_t(2a_i)=-B_i+\frac{3(12A_i^2-g_2)A_i}{2B_i}-\frac{(12A_i^2-g_2)^3}{32B_i^3} \hfill
\end{gathered}
\end{equation}
While solving the algebraic system we can simplify computations if we find the values $\zeta(a_i-a_j)$, $\wp(a_i-a_j)$, $\wp_t(a_i-a_j)$ first and only then use addition formulae \eqref{Addition_formulae} or \eqref{Addition_formulae_double}.

With the help of our method one can construct any elliptic
solution of equation \eqref{EQN}. Note that if equation \eqref{EQN}
possesses only $N$ distinct Laurent series in a neighborhood of
poles, then the orders of its elliptic solutions are not more than
$\sum_{i=1}^{N}p_i$, where $p_i$ ($i=1$, $\ldots$, $N$) are the
orders of poles given by the local singularity analysis \cite{Demina01, Demina02}. Thus we see that our
approach may be used, if one needs to classify families of elliptic
solutions satisfying equation \eqref{EQN}. Further, we mention that in the case $g_2^3-27g_3^2=0$ the elliptic function $\wp(t)$
degenerates and consequently elliptic solution
\eqref{Ex_Sol_EllipticK_AB} degenerates.

\section{Local singularity analysis for the H\'{e}non -- Heiles model} \label{HH_Loca_Analysis}

The method presented in section \ref{Method applied_HH} can be easily extended to a system of autonomous nonlinear ordinary differential
equations. Nevertheless in this article we shall consider the forth--order equation
\begin{equation}
\label{HH_eqn}
y_{tttt}-2(\sigma-4)yy_{tt}-2(\sigma+1)y_t^2-\frac{20}{3}\sigma y^3+(4\alpha_1+\alpha_2)y_{tt}+(6\alpha_2-4\sigma\alpha_1)y^2+4\alpha_1\alpha_2y+4H=0
\end{equation}
satisfied by the function $y(t)$. An advantage of this approach lies in the fact that the only function supposed to be elliptic is $y(t)$, while  this may not be true for $x(t)$. Equation \eqref{HH_eqn} can be obtained in the following way. We differentiate the second relation in \eqref{HH_equations_of_motion} twice and use expression \eqref{HH_Hamiltonian} and the first relation in \eqref{HH_equations_of_motion} to eliminate $x^2$ and $x_{tt}$. Note that $H$ in \eqref{HH_eqn} is the energy of the system. Further, let us suppose that the variable $t$ is complex and~$\sigma \neq0$.

We use the Painlev\'{e} methods to obtain the Laurent series satisfying equation \eqref{HH_eqn}. The dominant behaviors and the Fuchs indices are the following
\begin{equation}\begin{gathered}
\label{HH_eqn_dominant_behavior}
y^{(1)}(t)=-\frac{3}{t^2},\quad j=-1,\quad  10,\quad  \frac{5}{2}\pm \frac{1}{2}\sqrt{-23-24\sigma},\hfill\\
y^{(2)}(t)=\frac{6}{\sigma t^2},\quad j=-1,\quad  5,\quad  5\pm \sqrt{1-\frac{48}{\sigma}}.\hfill
\end{gathered}\end{equation}
Note that we omit the arbitrary constant $t_0$. Consequently, equation \eqref{HH_eqn} may admit two families of Laurent series:
\begin{equation}\begin{gathered}
\label{HH_eqn_LS}
y^{(1)}(t)=-\frac{3}{t^2}+\frac{c^{(1)}_{-1}}{t}+\sum_{k=0}^{\infty}c^{(1)}_{k}t^k,\hfill\\
y^{(2)}(t)=\frac{6}{\sigma t^2}+\frac{c^{(2)}_{-1}}{t}+\sum_{k=0}^{\infty}c^{(2)}_{k}t^k.\hfill
\end{gathered}\end{equation}
If the parameter $\sigma$ is such that the first family does not have positive integer Fuchs indices with the exception of $j=10$, then the series $y^{(1)}(t)$ in fact exists and possesses one arbitrary coefficient $c_8^{(1)}$ (in addition to the arbitrary constant $t_0$). A similar statement is true for the second family. The Laurent series $y^{(2)}(t)$ with arbitrary coefficient $c_3^{(2)}$ exists provided that the parameter $\sigma$ is chosen in such a way that there are no  positive integer Fuchs indices with the exception of $j=5$. In other cases additional arbitrary coefficients may enter the Laurent series. Thus such cases should be considered separately.

We would like to note that the equalities $c^{(1)}_{-1}=0$, $c^{(2)}_{-1}=0$ are valid unless the corresponding series posses the Fuchs index $j=1$. For the series  $y^{(1)}(t)$ this situation takes place in the case $\sigma=-4/3$ and for   the series  $y^{(2)}(t)$ such a situation occurs if $\sigma=-16/5$.

Let us take $\sigma=-4/3$, then the Laurent series $y^{(1)}(t)$ possesses two arbitrary coefficients $c_2^{(1)}$, $c_8^{(1)}$ corresponding to the Fuchs indices $j=4$, $j=10$ accordingly. However this series satisfies equation \eqref{HH_eqn} under the condition
\begin{equation}
\label{HH_minus4_3_Cond}
c_{-1}^4+\frac{2}{11}(4\alpha_1-3\alpha_2)c_{-1}^2+\frac{72}{1925}(2\alpha_1-\alpha_2)(\alpha_1-\alpha_2)=0,\quad c_{-1}\equiv c^{(1)}_{-1}.
\end{equation}
Analogously, setting $\sigma=-16/5$, we obtain the Laurent series $y^{(2)}(t)$, which may have two arbitrary coefficients $c_3^{(2)}$, $c_7^{(2)}$ associated with the Fuchs indices $j=5$, $j=9$ accordingly. This series satisfies equation \eqref{HH_eqn} provided that one of the following conditions
\begin{equation}
\label{HH_minus16_5_Cond}
\begin{gathered}
(i) \quad c_{-1}=0,\quad  c_{3}(16\alpha_1-\alpha_2)(16\alpha_1-9\alpha_2)=0,\quad c_{-1}\equiv c^{(2)}_{-1},\quad  c_{3}\equiv c^{(2)}_{3};\hfill\\
(ii) \quad c_{-1}^4+\frac{675\left(16\alpha_1-5\alpha_2\right)}{47872}c_{-1}^2+\frac{10125(16\alpha_1-\alpha_2)(16\alpha_1-9\alpha_2)}{343146496}=0,\quad c_{-1}\equiv c^{(2)}_{-1}.\hfill
\end{gathered}
\end{equation}
holds. From \eqref{HH_minus16_5_Cond} we see that the coefficient $c_7^{(2)}$ is always arbitrary and the coefficient $c_3^{(2)}$ is arbitrary either in the case (ii) or in the case (ii) with $\alpha_2=16\alpha_1$ or $\alpha_2=16\alpha_1/9$.

Further let us consider the values of $\sigma$, which correspond to integrable cases \eqref{HH_integrable cases}. We begin with $\sigma=-1$. The Laurent series $y^{(1)}(t)$ exists only in the case $\alpha_2=\alpha_1$ and possesses three arbitrary coefficients   $c_0^{(1)}$, $c_1^{(1)}$,  $c_8^{(1)}$  related to the Fuchs indices $j=2$, $j=3$, $j=10$ accordingly. The Laurent series $y^{(2)}(t)$ satisfies equation \eqref{HH_eqn} under the condition
\begin{equation}
\label{HH_minus1_Cond}
\begin{gathered}
c_{3}(\alpha_1-\alpha_2)=0,\quad c_{3}\equiv c^{(2)}_{3}
\end{gathered}
\end{equation}
and possesses at least one arbitrary coefficient  $c_{10}^{(2)}$. In the integrable case $\alpha_2=\alpha_2$ the series $y^{(2)}(t)$ admits another arbitrary coefficient $c_3^{(2)}$.  Interestingly that the series $y^{(2)}(t)$ may exist in the case $\alpha_2\neq \alpha_2$, but with  the loss of the arbitrary coefficient $c_3^{(2)}$, which is zero.

In the case $\sigma=-6$ the Laurent series $y^{(1)}(t)$, $y^{(2)}(t)$ satisfy equation \eqref{HH_eqn} at any values of the parameters $\alpha_1$, $\alpha_2$. The series $y^{(2)}(t)$ possesses three arbitrary coefficients: $c_{0}^{(2)}$, $c_{3}^{(2)}$, $c_{6}^{(2)}$, while the series $y^{(1)}(t)$ only two: $c_{6}^{(1)}$, $c_{8}^{(1)}$.

Finally in the case $\sigma=-16$ equation \eqref{HH_eqn} admits the Laurent series $y^{(1)}(t)$ whenever one of the following conditions hold
\begin{equation}
\label{HH_minus16_Cond1}
\begin{gathered}
(i)\quad \alpha_2=16\alpha_1, \hfill \\
(ii)\quad c_{8}=\frac{[14\alpha_1\alpha_2-112\alpha_1^2-\alpha_2^2]\lambda}{47567520},\quad  c_{8}\equiv c^{(1)}_{8},\hfill \\
\lambda =\frac {7658}{22275}\alpha_1^3-\frac {3829}{59400}\alpha_1^2\alpha_2-\frac {1057}{118800}\alpha_1\alpha_2^2+
\frac {71}{178200}\alpha_2^3+H
\end{gathered}
\end{equation}
In the case $(i)$ the coefficients $c_{8}^{(1)}$, $c_{10}^{(1)}$ are arbitrary, while in the case $(ii)$ there exists only one arbitrary coefficient: $c_{10}^{(1)}$. The  Laurent series $y^{(2)}(t)$ satisfies equation \eqref{HH_eqn} under the conditions
\begin{equation}
\label{HH_minus16_Cond2}
\begin{gathered}
(i)\quad \alpha_2=16\alpha_1, \hfill \\
(ii)\quad c_{1}^{(2)}=0,\quad c_{3}^{(2)}=0 \hfill
\end{gathered}
\end{equation}
and possesses three arbitrary constants $c_{1}^{(2)}$, $c_{3}^{(2)}$, $c_{5}^{(2)}$ in the case $(i)$  and only one $c_{5}^{(2)}$ in the case $(ii)$ . Again we note that one can construct the Laurent series in the situation $\alpha_2 \neq 16\alpha_1$.

In addition let us note that in the case $\sigma=-2$ the series $y^{(1)}(t)$, $y^{(2)}(t)$ coincide. Such a series possesses an arbitrary coefficient $c_8$ and exists under the condition $c_3=0$.

The local analysis shows that elliptic solutions of equation \eqref{HH_eqn} possess even orders. In the next sections we  shall classify second--order, fourth--order, and sixth--order elliptic solutions.

\section{Second--order elliptic solutions of equation \eqref{HH_eqn}}\label{HH_SO_ES}

In this section let us construct second--order elliptic solutions of equation \eqref{HH_eqn}. Using results of section \ref{HH_Loca_Analysis} and relation \eqref{Ex_Sol_EllipticK}, we find the general expression for second--order elliptic functions that satisfy equation \eqref{HH_eqn}:
\begin{equation}
\label{HH_SO_General}
y(t)=c_{-2}\wp(t,g_2,g_3)+h_0.
\end{equation}
Again we omit the arbitrary constant $t_0$. Constructing the Laurent series for this function in a neighborhood of its pole $t=0$, we obtain
\begin{equation}
\begin{gathered}
\label{HH_SO_General_LS}y(t)=\frac{c_{-2}}{t^2}+h_0+\frac{g_2}{20}c_{-2}t^2+\frac{g_3}{28}c_{-2}t^4+\ldots.
\end{gathered}
\end{equation}

According to the method of section  \ref{Method applied_HH} we should either compare the coefficient of the series found by the methods of local analysis with coefficients of series \eqref{HH_SO_General_LS} or substitute series  \eqref{HH_SO_General_LS} into the original equation and set to zero the coefficients at negative and zero powers of $t$. Since series \eqref{HH_eqn_LS} may possess arbitrary coefficients, we shall use the second alternative.

Let us present our results for the case $c_{-2}=-3$. We have found five families of elliptic solutions with the local behavior $y^{(1)}(t)$.  The parameters $h_0$, $g_2$, $g_3$ for the first family of elliptic solutions are the following
\begin{equation}
\begin{gathered}
\label{HH_SO_Sol1}h_0=\frac{\alpha_2-(\sigma+2)\alpha_1}{4(\sigma+1)},\quad  g_2=\frac{(\sigma+2)(3\sigma-2)(\sigma\alpha_1+2\alpha_2)\alpha_1+(7\sigma+12)\alpha_2^2}{12(\sigma+1)^2(3\sigma+4)},\hfill\\
g_3=\frac{ \left( \sigma+2 \right)\alpha_1\alpha_2\delta_1 +\delta_2
-48H\left( 3\,\sigma+4 \right)  \left( \sigma+1 \right) ^{3} }{216(\sigma+1)^3(3\sigma+4)(\sigma+3)},\hfill\\
\end{gathered}
\end{equation}
where we have used the designations
\begin{equation}
\begin{gathered}
\label{HH_SO_Sol1_delta}
\delta_1= 3\left( 3{\sigma}^{3}+7\,{\sigma}^{2}+28
\,\sigma+4 \right)\alpha_1-3(7\sigma^2+17\sigma-10)\alpha_2,\hfill\\
\delta_2=\left( \sigma+2 \right) \left( 3{\sigma}^{3}+7{\sigma}^{
2}+28\,\sigma+4 \right)\sigma
\alpha_1^3- \left( 3\sigma+8 \right)  \left( 5\sigma+9 \right) \alpha_2^3 \hfill
\end{gathered}
\end{equation}
The first family exists provided that $\sigma\neq-1$, $\sigma\neq-3$, $\sigma\neq-4/3$ and appeared in article~\cite{Vernov01}.

The parameters $g_2$, $g_3$  for the second family are given by
\begin{equation}
\begin{gathered}
\label{HH_SO_Sol2}g_2=\frac{4\left\{5h_0^2+5\alpha_1h_0+\alpha_1^2\right\}}{3},\, g_3=-\frac{320h_0^{3}+480\alpha_1h_0^2+222\alpha_1^2h_0+30\alpha_1^3+12H}{108}
\end{gathered}
\end{equation}
This family satisfies equation \eqref{HH_eqn} under the conditions $\sigma=-1$, $\alpha_2=\alpha_1$ and possess the arbitrary constant $h_0$. In the case $\sigma=-3$ we obtain the third family with the parameters
\begin{equation}
\begin{gathered}
\label{HH_SO_Sol3}h_0=-\frac{\alpha_1+\alpha_2}{8},\quad g_2=\frac{9\alpha_2^{2}-22\alpha_2\alpha_1+33\alpha_1^2}{240}.
\end{gathered}
\end{equation}
This family exists under the condition
\begin{equation}
\begin{gathered}
\label{HH_SO_Sol3_H}H=-\frac{(\alpha_2-\alpha_1)(\alpha_2-7\alpha_1)(\alpha_2+7\alpha_1)}{320}.
\end{gathered}
\end{equation}
and possesses the arbitrary constant $g_3$. In the case $\sigma=-4/3$ we obtain two families (the fourth family and the fifth family in our numeration) of elliptic solutions. The fourth family  satisfies equation \eqref{HH_eqn} under the conditions $\alpha_2=\alpha_1$ (see \eqref{HH_minus4_3_Cond} with $c_1=0$). The parameters $h_0$, $g_3$ are the  following
\begin{equation}
\begin{gathered}
\label{HH_SO_Sol4}h_0=-\frac{\alpha_1}{4},\quad g_3={\frac {31}{2160}}\alpha_1^3-{\frac {7}{60}}\alpha_1g_{{2}}-\frac2{15}H.
\end{gathered}
\end{equation}
The fifth family solve equation \eqref{HH_eqn} provided  that $\alpha_2=2\alpha_1$ (see \eqref{HH_minus4_3_Cond} with $c_1=0$). The parameters $h_0$, $g_3$ are given by
\begin{equation}
\begin{gathered}
\label{HH_SO_Sol5}h_0=-\alpha_1,\quad g_3=\frac {7}{30}\alpha_1g_{2}-{\frac {2}{135}}\alpha_1^3-\frac2{15}H.
\end{gathered}
\end{equation}
 For the last two families the parameter $g_2$ is arbitrary.
%\begin{equation}
%\begin{gathered}
%\label{HH_SO_Sol4}h_0=\frac{2\alpha_1-3\alpha_2}{4},\, g_3=\frac{\left( 3\alpha_2-2\alpha_1 \right)  \left( 9\alpha_2^2-66\alpha_2\alpha_1+88\alpha_1^2\right)
%+252\left( 3\alpha_2 -4\alpha_1\right)g_2-288H}{2160}.
%\end{gathered}
%\end{equation}

Further we proceed to the case $c_{-2}=6/\sigma$. We have obtained two families of elliptic solutions with the local behavior $y^{(2)}(t)$. The parameters $h_0$, $g_2$, $g_3$ for the first family are
\begin{equation}
\begin{gathered}
\label{HH_SO_Sol6}h_0=\frac{\alpha_2}{2\sigma},\quad g_2=\frac{\alpha_2^2}{12},\quad g_3=\frac{\alpha_2^3}{216}-\frac{\sigma^2H}{18}.
\end{gathered}
\end{equation}
If $\sigma=-6$ we can construct two--parametric elliptic solution, which generalizes  \eqref{HH_SO_Sol6}. Thus, the second family exists in the case $\sigma=-6$ and possesses the arbitrary parameter $h_0$. The other parameters are given by
\begin{equation}
\begin{gathered}
\label{HH_SO_Sol7}g_2=-6h_0\{10h_0+4\alpha_1+\alpha_2\}-2\alpha_1\alpha_2, \quad g_3=14h_0^2\{ 20h_0+3\alpha_2+12\alpha_1\} \\
+\frac{h_0}2\{4\alpha_1+3\alpha_2\} \{12\alpha_1+\alpha_2\}+\frac{\alpha_1\alpha_2}{2}\{4\alpha_1+\alpha_2\}
-2H. \hfill
\end{gathered}
\end{equation}
%\begin{equation}
%\begin{gathered}
%\label{HH_SO_Sol7}g_2=-60h_0^2-2\alpha_1\alpha_2-6\alpha_2h_0-24\alpha_1h_0, \hfill\\ %g_3=280h_0^3+20\alpha_1\alpha_2h_0+42\alpha_2h_0^2+168\alpha_1h_0^2+2\alpha_1^2\alpha_2+24\alpha_1^2h_0+\frac12\alpha_1\alpha_2^2+\frac32\alpha_2^2h_0-2H.
%\end{gathered}
%\end{equation}

The families of second--order elliptic solutions with $\sigma=-3$, $\sigma=-3/4$ seem to be new, while the families with $\sigma=-1$, $\sigma=-6$ are reductions of known general solutions in the integrable cases. Concluding this section we would like to note that elliptic solutions we have found degenerate if the following condition is valid: $g_2^3-27g_3^2=0$.

\section{Fourth--order elliptic solutions of equation \eqref{HH_eqn}}\label{HH_FO_ES}

In this section our aim is to find fourth--order elliptic solutions of equation \eqref{HH_eqn}. Using formula \eqref{Ex_Sol_EllipticK} and the results of asymptotic analysis, we see that if such an elliptic solution exists, then it necessary possesses two double poles inside a parallelogram of periods. Without loss of generality let us omit the arbitrary constant $t_0$ and suppose that these poles are $t=0$, $t=a$. There are two possibilities. Equation \eqref{HH_eqn} may admit   fourth--order  elliptic solutions possessing simultaneously poles with the Laurent series given by $y^{(1)}(t)$, $y^{(2)}(t-a)$.  In addition there may exist fourth--order elliptic solutions possessing  poles with the Laurent series of only one type. The latter situation may take place only if the series corresponding to the poles $t=0$, $t=a$ are in fact distinct. In the first situation we get the following expressions for the fourth--order elliptic solutions
\begin{equation}
\begin{gathered}
\label{HH_FO_General1}y(t)=c_{-2}^{(1)}\wp(t,g_2,g_3)+c_{-2}^{(2)}\wp(t-a,g_2,g_3)+\tilde{h}_0,
\end{gathered}
\end{equation}
where $\tilde{h}_0=h_0+c_{-2}^{(2)}\wp(a,g_2,g_3)$. The Laurent series of this function in a neighborhood of the poles $t=0$, $t=a$  are of the form
\begin{equation}
\begin{gathered}
\label{HH_FO_General_LS1}y(t)=\frac{c_{-2}^{(1)}}{t^2}+h_0+2Ac_{-2}^{(2)}-Bc_{-2}^{(2)}t+\left\{\left(3A^2-\frac{g_2}{4}\right)c_{-2}^{(2)}+\frac{g_2}{20}c_{-2}^{(1)}\right\}t^2+\ldots\hfill \\
y(t)=\frac{c_{-2}^{(2)}}{(t-a)^2}+h_0+2Ac_{-2}^{(1)}+Bc_{-2}^{(1)}(t-a)+\left\{\left(3A^2-\frac{g_2}{4}\right)c_{-2}^{(1)}+\frac{g_2}{20}c_{-2}^{(2)}\right\}(t-a)^2+\ldots
\end{gathered}
\end{equation}
Note that throughout this section we use notation $A\stackrel{def}{=}\wp(a,g_2,g_3)$, $B\stackrel{def}{=}\wp_t(a,g_2,g_3)$. In the second situation the general expression for the fourth--order elliptic solutions can be written~as
\begin{equation}
\begin{gathered}
\label{HH_FO_General2}y(t)=c_{-2}\wp(t,g_2,g_3)+b\zeta(t,g_2,g_3)+c_{-2}\wp(t-a,g_2,g_3)-b\zeta(t-a,g_2,g_3)+\tilde{h}_0,
\end{gathered}
\end{equation}
where $\tilde{h}_0=h_0+c_{-2}\wp(a,g_2,g_3)-b\zeta(a,g_2,g_3)$, $c_{-2}=c_{-2}^{(1)}$ or $c_{-2}=c_{-2}^{(2)}$. Finding the Laurent series of the function \eqref{HH_FO_General2} in a neighborhood of the poles $t=0$, $t=a$, we obtain the series
\begin{equation}
\begin{gathered}
\label{HH_FO_General_LS2}y(t)=\frac{c_{-2}}{t^2}+\frac{b}{t}+h_0+2Ac_{-2}+\{bA-Bc_{-2}\}t+\left\{\left(3A^2-\frac{g_2}{5}\right)c_{-2}^{(2)}-\frac{b}{2}B\right\}t^2+\ldots\hfill \\
y(t)=\frac{c_{-2}}{(t-a)^2}-\frac{b}{t-a}+h_0+2Ac_{-2}-\{bA-Bc_{-2}\}(t-a)\hfill \\
+\left\{\left(3A^2-\frac{g_2}{5}\right)c_{-2}-\frac{b}{2}B\right\}(t-a)^2+\ldots \hfill
\end{gathered}
\end{equation}
We see that the series given in \eqref{HH_FO_General_LS2} are distinct in the following cases: $b\neq 0$ or $b=0$, $B\neq0$. Thus we should consider the values of $\sigma$ giving the Fuchs indices $j=1$ or (and)~$j=3$ for at least one of the series $y^{(1)}(t)$, $y^{(2)}(t)$ (see table \ref{T:HH_FO2}).
\begin{table}[h]%[h]
    \caption{Values of $\sigma$ with the Fuchs indices $j=1$ or (and)~$j=3$.} \label{T:HH_FO2}
    \center
       \begin{tabular}[pos]{|c|c|c|}
                \hline
                Values of $\sigma$ & Fuchs indices & Type of series\\
                \hline
$-\frac43$ & $ -1,1,4,10$ & $y^{(1)}(t)$\\
\hline
$-1$ & $ -1,2,3,10$ & $y^{(1)}(t)$\\
\hline
$-\frac{16}{5}$ & $ -1,1,5,9$ & $y^{(2)}(t)$\\
\hline
$-16$ & $ -1,3,5,7$ & $y^{(2)}(t)$\\
\hline
        \end{tabular}
\end{table}

Let us present our results. In order to find elliptic solutions of the form \eqref{HH_FO_General1} or \eqref{HH_FO_General2} we calculate five coefficients $c_{k}$, $k=0$, $\ldots$, $4$ in each of the series \eqref{HH_FO_General_LS1} or \eqref{HH_FO_General_LS2}. Substituting these series into equation \eqref{HH_eqn}, we get $10$ nontrivial algebraic equations. Further we solve the algebraic system with an additional equation $B^2=4A^3-g_2A-g_3$. There exists a number of solutions \eqref{HH_FO_General1} corresponding to degenerate elliptic solutions. We do not present these solutions here. The only case, which leads to non--degenerate elliptic solutions of the form \eqref{HH_FO_General1}, is $\sigma=-6$. This family of elliptic solutions can be written as
\begin{equation}
\begin{gathered}
\label{HH_FO_Sol1}y(t)=-\frac{1}{4}\left[\frac{\wp_t(t;g_2,g_3)}{\wp(t;g_2,g_3)-A}\right]^2-2\wp(t;g_2,g_3)+2A-\frac{\alpha_2}{20}-\frac{\alpha_1}{5},
\end{gathered}
\end{equation}
where the invariants $g_2$, $g_3$ are given by
\begin{equation}
\begin{gathered}
\label{HH_FO_Sol1_g}g_2=30A^2-\frac {3}{280}\alpha_2^2+\frac {2}{35}\alpha_2\alpha_1-\frac {6}{35}\alpha_1^2,\, g_3 = -\frac {A}{280}\left( 7280A^2-3\alpha_2^2+16\alpha_2\alpha_1-48\alpha_1^2 \right)
\end{gathered}
\end{equation}
and the parameter $A$ is one of the roots of the following cubic equation
\begin{equation}
\begin{gathered}
\label{HH_FO_Sol1_A}A^3+\left(\frac{\alpha_2\alpha_1}{357}-\frac{\alpha_2^2}{1904}-\frac{\alpha_1^2}{119}\right) A+\frac{H}{918}+\frac{\alpha_2^3}{367200}-\frac{\alpha_2\alpha_1^2}{11475}-\frac{\alpha_2^2\alpha_1}{45900}+\frac{2\alpha_1^3}{11475}=0
\end{gathered}
\end{equation}
Note that we have used an addition formula for the Weierstrass elliptic function (see \eqref{Ex_Sol_EllipticK_AB}) in order to rewrite expression \eqref{HH_FO_General1} in the form \eqref{HH_FO_Sol1}.

\begin{table}[t]%[h]
    \caption{Coefficients of relations \eqref{HH_FO_Sol2_g}, \eqref{HH_FO_Sol23_A} in the case $\sigma=-4/3$.} \label{T:HH_2poles2}
    \center
       \begin{tabular}[pos]{|l|}
                 \hline
                \\
                 $\gamma_1=-\frac {188}{77}\alpha_1^2+\frac {282}{77}\alpha_2\alpha_1-\frac {3923}{2772}\alpha_2^2$\\
                \\
                $\gamma_2=-\frac {70}{9}A^2-\left\{\frac {6797}{891}\alpha_1+\frac {6797}{1188}\alpha_2\right\}A+\frac{604523}{294030}\alpha_2\alpha_1-\frac {310253}{392040}\alpha_2^2-\frac {604523}{441045}\alpha_1^2 $\\
                \\
                $ \gamma_3=-\frac {4}{135}H-\frac {22552}{27225}\alpha_1^3+\frac {239873}{653400}\alpha_2^3-\frac {1391287}{980100}\alpha_2^2\alpha_1+\frac {5638}{3025}\alpha_1^2\alpha_2$\\
                \\
                \hline
                \\
                 $\mu_2=\frac {11}{24}\alpha_2-\frac {11}{18}\alpha_1 -\frac {29}{24}b^2$\\
                 \\
                 $\mu_1=\frac {2581}{14256}\{3\alpha_2-4\alpha_1\}b^2-\frac {22522}
{51975}\alpha_1^2+\frac {11261}{17325}\alpha_2\alpha_1-\frac {11507}{46200}\alpha_2^2$\\
                 \\
                 $ \mu_0=\left\{\frac {10961063}{82328400}\alpha_2\alpha_1-\frac {10961063}{123492600}\alpha_1^2-\frac {22367087}{439084800}\alpha_2^2\right\}b^2-\frac {1439023}{15681600}\alpha_2^2\alpha_1
+\frac {247217}{10454400}\alpha_2^3$\\
\\
$\qquad -\frac {2921}{54450}\alpha_1^3+
\frac {2921}{24200}\alpha_2\alpha_1^2-\frac {H}{540}$\\
\\
                \hline
        \end{tabular}
\end{table}

Now let us find fourth--order elliptic solutions of the form \eqref{HH_FO_General2}. We begin with the case $\sigma=-4/3$. From table \ref{T:HH_FO2} it follows that we should take $c_{-2}=-3$ in expression \eqref{HH_FO_General2}. The local singularity analysis (see \eqref{HH_minus4_3_Cond}) shows that the parameter $b$ satisfies the relation
 \begin{equation}
\label{HH_FO_Sol2_b}
b^4+\frac{2}{11}(4\alpha_1-3\alpha_2)b^2+\frac{72}{1925}(2\alpha_1-\alpha_2)(\alpha_1-\alpha_2)=0.
\end{equation}
Solving $11$ algebraic equations, we obtain the following family of elliptic solutions
\begin{equation}
\begin{gathered}
\label{HH_FO_Sol2}y(t)=-\frac34\left[\frac{\wp_t(t;g_2,g_3)+B}{\wp(t;g_2,g_3)-A}\right]^2-\frac{b}{2}\left[\frac{\wp_t(t;g_2,g_3)+B}{\wp(t;g_2,g_3)-A}\right]+h_0,
\end{gathered}
\end{equation}
where the parameters $h_0$, $B$, $g_2$, $g_3$ are given by
\begin{equation}
\begin{gathered}
\label{HH_FO_Sol2_g}h_0=6A-\frac34\alpha_2+\frac12 \alpha_1+\frac {29}{24}b^2,\quad B=-\frac{b}{36} \left(12A -34b^2+15\alpha_2-20\alpha_1\right),\hfill \\
g_2=20A^2+\left[\frac {3605}{1188}\alpha_2-\frac {3605}{891}\alpha_1-\frac {35}{3}A \right]b^2+5\left[\alpha_2-\frac43\alpha_1\right]A+\gamma_1,\hfill \\
g_3=\gamma_2b^2-\frac{A}{1188}\left(924\{4\alpha_1-3\alpha_2\}A+3081\alpha_2^2-8088\alpha_2\alpha_1+5392\alpha_1^2\right)+\gamma_3
\end{gathered}
\end{equation}
The parameter $A$ is one of the roots of the cubic equation
\begin{equation}
\begin{gathered}
\label{HH_FO_Sol23_A}A^3+\mu_2A^2+\mu_1A+\mu_0=0,
\end{gathered}
\end{equation}
while the parameter $b \neq 0$ is a root of equation \eqref{HH_FO_Sol2_b}. The coefficients $\gamma_1$, $\gamma_2$, $\mu_0$, $\mu_1$, $\mu_2$ in expressions \eqref{HH_FO_Sol2_g}, \eqref{HH_FO_Sol23_A} are presented in table \ref{T:HH_2poles2}.
%\begin{equation}
%\begin{gathered}
%\label{HH_FO_Sol2_gamma}\gamma_1=-\frac {188}{77}\alpha_1^2+\frac {282}{77}\alpha_2\alpha_1-\frac {3923}{2772}\alpha_2^2,\hfill \\
%\gamma_2=-\frac {70}{9}A^2-\left\{\frac {6797}{891}\alpha_1+\frac {6797}{1188}\alpha_2\right\}A+\frac{604523}{294030}\alpha_2\alpha_1-\frac {310253}{392040}\alpha_2^2-\frac %{604523}{441045}\alpha_1^2,\\
%\gamma_3=-\frac {4}{135}H-\frac {22552}{27225}\alpha_1^3+\frac {239873}{653400}\alpha_2^3-\frac {1391287}{980100}\alpha_2^2\alpha_1+\frac {5638}{3025}\alpha_1^2\alpha_2 \hfill
%\end{gathered}
%\end{equation}

%\begin{equation}
%\begin{gathered}
%\label{HH_FO_Sol2_A}A^3+ \left(\frac {11}{24}\alpha_2-\frac {11}{18}\alpha_1 -\frac {29}{24}b^2\right)A^2+ \left( \frac {2581}{14256}\{3\alpha_2-4\alpha_1\}b^2-\frac {22522}
%{51975}\alpha_1^2+\frac {11261}{17325}\alpha_2\alpha_1-\frac {11507}{46200}\alpha_2^2 \right)A+\gamma_3 \hfill \\
%\gamma_3=\left\{\frac {10961063}{82328400}\alpha_2\alpha_1-\frac {10961063}{123492600}\alpha_1^2-\frac {22367087}{439084800}\alpha_2^2\right\}b^2-\frac {1439023}{15681600}\alpha_2^2\alpha_1\\
%+\frac {247217}{10454400}\alpha_2^3-\frac {2921}{54450}\alpha_1^3+
%\frac {2921}{24200}\alpha_2\alpha_1^2-\frac {H}{540}
%\end{gathered}
%\end{equation}

Further we proceed to the case $\sigma=-16/5$. Using table \ref{T:HH_FO2}, we set $c_{-2}=-15/8$ in expression \eqref{HH_FO_General2}. While solving the algebraic system, we obtain that from $b=0$ it follows $B=0$, Consequently, we suppose that $b\neq 0$ and consider the case (ii) in relation \eqref{HH_minus16_5_Cond} with $c_{-1}=b$. We find the following family of elliptic solutions
\begin{equation}
\begin{gathered}
\label{HH_FO_Sol3}y(t)=-\frac{15}{32}\left[\frac{\wp_t(t;g_2,g_3)+B}{\wp(t;g_2,g_3)-A}\right]^2-\frac{b}{2}\left[\frac{\wp_t(t;g_2,g_3)+B}{\wp(t;g_2,g_3)-A}\right]+h_0,
\end{gathered}
\end{equation}
where the parameters $h_0$, $B$, $g_2$, $g_3$ take the form
\begin{equation}
\begin{gathered}
\label{HH_FO_Sol3_g}h_0=\frac{62}{45}b^2-\frac {5}{32}\alpha_2+\frac{15}{4}A,\quad B=\left(\frac29\alpha_2+\frac {5056}{3375}b^2-\frac{5}{72}\alpha_2-\frac {8}{15}A\right)b, \hfill \\
g_2= \left( \frac {1697}{5049}\alpha_2-\frac {27152}{25245}\alpha_1-\frac {32}{45}A \right)b^2+15A^2+\delta_1,\hfill \\
g_3=\delta_2b^2+\left(\frac {4135}{8976}\alpha_2\alpha_1-\frac {17915}{287232}\alpha_2^2-\frac {827}{1122}\alpha_1^2\right)A+\delta_3 \hfill
\end{gathered}
\end{equation}
and the parameter $A$ is a root of the cubic equation \eqref{HH_FO_Sol23_A} with the coefficients $\mu_0$, $\mu_1$, $\mu_2$ given in table \ref{T:HH_2poles3}. The parameter $b \neq 0$ is a root of the equation
\begin{equation}
\label{HH_FO_Sol3_b}
\begin{gathered}
 b^4+\frac{675\left(16\alpha_1-5\alpha_2\right)}{47872}b^2+\frac{10125(16\alpha_1-\alpha_2)(16\alpha_1-9\alpha_2)}{343146496}=0
\end{gathered}
\end{equation}
and the coefficients $\mu_0$, $\mu_1$, $\mu_2$ are presented in table \ref{T:HH_2poles3}.

\begin{table}[t]%[h]
    \caption{Coefficients of relations  in the case $\sigma=-16/5$.} \label{T:HH_2poles3}
    \center
       \begin{tabular}[pos]{|l|}
                 \hline
                \\
                 $\delta_1=-\frac {3433}{15708}\alpha_1^2+\frac {17165}{125664}\alpha_2\alpha_1-\frac {114673}{4021248}\alpha_2^2$\\
                \\
                $\delta_2=-\frac {896}{45}A^2+ \left\{\frac {14693}{8415}\alpha_2-\frac {235088}{42075}\alpha_1 \right\}A-\frac {80008}{147015}\alpha_1^2+\frac {10001}{29403}\alpha_2\alpha_1-\frac {455047}{7840800}\alpha_2^2$\\
                \\
                $ \delta_3=-\frac {307}{2420}\alpha_1^3+\frac{921}{7744}\alpha_1^2\alpha_2+\frac {159785}{53526528}\alpha_2^3-\frac {18113}{619520}\alpha_2^2\alpha_1-\frac {44}{225}H$\\
                \\
                \hline
                \\
                 $\mu_2=-\frac{416}{225}b^2$\\
                 \\
                 $\mu_1=\frac {1}{225}\left\{\frac {23450}{561}\alpha_2+\frac{75040}{561}\alpha_1 \right\}b^2+\frac {6737}{125664}\alpha_2\alpha_1-\frac {165353}{20106240}\alpha_2^2-\frac{6737}{78540}\alpha_1^2$\\
                 \\
                 $ \mu_0=\frac {1}{225}\left\{ \frac {46282237}{6609141}\alpha_2\alpha_1+\frac {370257896}{33045705}\alpha_1^2+\frac
{413300471}{352487520}\alpha_2^2\right\}b^2+
\frac {24427}{2238016}\alpha_1^2\alpha_2-\frac {1441193}{537123840}\alpha_1\alpha_2^2$\\
\\
$\qquad -\frac {24427}{2098140}\alpha_1^3+\frac {4216603}{15469166592}\alpha_2^3-\frac {4}{225}H$\\
\\
                \hline
        \end{tabular}
\end{table}

Now let us consider the case $\sigma=-1$. Setting $c_{-2}=-3$ in expression \eqref{HH_FO_General2}, we see that the series $y^{(1)}(t)$ exists only in the case $\alpha_2=\alpha_1$. Since $b=0$, we have $B\neq 0$. Solving the algebraic system, we find the following family of elliptic solutions
\begin{equation}
\begin{gathered}
\label{HH_FO_Sol4}y(t)=-\frac{3}{4}\left\{\frac{36\wp_t(t;g_2,g_3)+\sqrt {5184A^3-108A\alpha_1^2-6\alpha_1^3+72H}}{36\left(\wp(t;g_2,g_3)-A\right)}\right\}^2+3A-\frac12 \alpha_1,
\end{gathered}
\end{equation}
where the invariant $g_2$, $g_3$ are given by
\begin{equation}
\begin{gathered}
\label{HH_FO_Sol4_g}g_2=\frac{1}{12}\alpha_1^2,\quad g_3=\frac{1}{216}\alpha_1^3-\frac{1}{18}H
\end{gathered}
\end{equation}
and the parameter $A$ is an arbitrary constant.

Further it remains to study the case $\sigma=-16$. We take $c_{-2}=-3/8$ in expression \eqref{HH_FO_General2} and come to a conclusion that $b=0$, $B\neq 0$. It follows from $B\neq 0$ that $c_{1}^{(2)}\neq 0$. Consequently, the series $y^{(2)}(t)$ exists only in the case $\alpha_2=16\alpha_1$ as given in \eqref{HH_minus16_Cond2}. We find the following family of elliptic solutions
\begin{equation}
\begin{gathered}
\label{HH_FO_Sol5}y(t)=-\frac{3}{32}\left\{\frac{9\wp_t(t;g_2,g_3)+\sqrt {36H-81A^3+108A\alpha_1^2-48\alpha_1^3}}{9\left(\wp(t;g_2,g_3)-A\right)}\right\}^2+\frac34A-\frac12 \alpha_1,
\end{gathered}
\end{equation}
where the invariant $g_2$, $g_3$ take the form
\begin{equation}
\begin{gathered}
\label{HH_FO_Sol5_g}g_2=15A^2-\frac{16}{3}\alpha_1^2,\quad g_3=4A\alpha_1^2-10A^3+\frac{16}{27}\alpha_1^3-\frac49H
\end{gathered}
\end{equation}
and the parameter $A$ is an arbitrary constant. Note that we may take any sign at the square root in expressions \eqref{HH_FO_Sol4}, \eqref{HH_FO_Sol5}.

We have finished the classification of fourth--order elliptic solutions of equation \eqref{HH_eqn}. The existence of fourth--order elliptic solutions with $\sigma=-4/3$, $\sigma=-16/5$ was claimed in \cite{Vernov02}, while in explicit form these solutions are presented for the first time. Elliptic solutions \eqref{HH_FO_Sol1}, \eqref{HH_FO_Sol4}, \eqref{HH_FO_Sol5} are reductions of hyperelliptic general solutions of equation \eqref{HH_eqn} with $\sigma=-6$, $\sigma=-1$, $\sigma=-16$. The degeneracy condition for the elliptic solutions constructed in this section is~$g_2^3-27g_3^2=0$.

\section{Sixth--order elliptic solutions of equation \eqref{HH_eqn}} \label{HH_SO_ES}

Let us classify sixth--order elliptic solutions of equation \eqref{HH_eqn}. It follows from formula \eqref{Ex_Sol_EllipticK} and the results of asymptotic analysis that such a solution possesses three double poles inside a parallelogram of periods. Without loss of generality we omit the arbitrary constant $t_0$ and suppose that these poles are $t=0$, $t=a_1$, $t=a_2$. Below we use the following notation $r_1=c_{-2}^{(1)}$, $r_2=c_{-2}^{(2)}$. There are four possibilities. Equation \eqref{HH_eqn} may admit   sixth--order  elliptic solutions possessing simultaneously poles with the Laurent series $y^{(2)}(t)$, $y^{(1)}(t-a_1)$, $y^{(1)}(t-a_2)$.  These elliptic solutions are given by the relation
\begin{equation}
\begin{gathered}
\label{HH_SO_General1}y(t)=r_2\wp(t)+r_1\wp(t-a_1)+b\zeta(t-a_1) +r_1\wp(t-a_2)-b\zeta(t-a_2)+\tilde{h}_0,
\end{gathered}
\end{equation}
where $\tilde{h}_0=h_0+r_1\wp(a_1)+r_2\wp(a_2)+b\zeta(a_1)-b\zeta(a_2)$ and  $b=0$ whenever $\sigma\neq -4/3$. Further equation \eqref{HH_eqn} may have  sixth--order elliptic solutions with the Laurent series $y^{(1)}(t)$, $y^{(2)}(t-a_1)$, $y^{(2)}(t-a_2)$ in a neighborhood the poles $t=0$, $t=a_1$, $t=a_2$. We obtain the following expression
\begin{equation}
\begin{gathered}
\label{HH_SO_General2}y(t)=r_1\wp(t)+r_2\wp(t-a_1)+b\zeta(t-a_1)+r_2\wp(t-a_2)-b\zeta(t-a_2)+\tilde{h}_0,
\end{gathered}
\end{equation}
where $\tilde{h}_0=h_0+r_2\wp(a_1)+r_2\wp(a_2)+b\zeta(a_1)-b\zeta(a_2)$ and $b=0$ whenever $\sigma\neq -16/5$. In addition there may exist sixth--order elliptic solutions of equation \eqref{HH_eqn} with the Laurent series in a neighborhood of the poles of only the first or the second type. Such solutions have the expression
\begin{equation}
\begin{gathered}
\label{HH_SO_General3_4}y(t)=r_i\wp(t)+b_1\zeta(t)+r_i\wp(t-a_1)+b_2\zeta(t-a_1)+r_i\wp(t-a_2)-(b_1+b_2)\zeta(t-a_2)\hfill \\
+\tilde{h}_0,\quad \tilde{h}_0=h_0+r_i\wp(a_1)+r_i\wp(a_2)+b_2\zeta(a_1)-(b_1+b_2)\zeta(a_2),
\end{gathered}
\end{equation}
where $i=1$ or $i=2$. In the case $i=1$ ($i=2$) we see that $b_1=0$ and $b_2=0$ whenever  $\sigma\neq -4/3$ ($\sigma\neq -16/5$).

The Laurent series of the functions \eqref{HH_SO_General1}, \eqref{HH_SO_General2}  in a neighborhood of the poles $t=0$, $t=a_1$, $t=a_2$  take the form
\begin{equation}
\begin{gathered}
\label{HH_SO_General_LS1_2}y(t)=\frac{r_i}{t^2}+2r_j(A_1+A_2)+h_0-\left\{r_j\left(B_1+B_2\right)+b\left(A_1-A_2\right)\right\}t+\ldots\hfill \\
y(t)=\frac{r_j}{(t-a_1)^2}+\frac{b}{t-a_1}+\nu_1+r_iA_1+\left\{r_iB_1+r_jP_{1,2}+bT_{1,2}\right\}(t-a_1)+\ldots \hfill \\
y(t)=\frac{r_j}{(t-a_2)^2}-\frac{b}{t-a_2}+\nu_1+r_iA_2+\left\{r_iB_2-r_jP_{1,2}-bT_{1,2}\right\}(t-a_2)+\ldots, \hfill
\end{gathered}
\end{equation}
where $\nu_1=r_j(A_1+A_2+T_{1,2})-bJ_{1,2}+h_0$. Along with this  we take $i=2$, $j=1$ for the function \eqref{HH_SO_General1} and  and we set $i=1$, $j=2$ for the function \eqref{HH_SO_General2}. Throughout this section we use notation $J_{1,2}\stackrel{def}{=}\zeta(a_1-a_2)-\zeta(a_1)+\zeta(a_2)$, $T_{1,2}\stackrel{def}{=}\wp(a_1-a_2)$, $P_{1,2}\stackrel{def}{=}\wp_t(a_1-a_2)$, $A_1\stackrel{def}{=}\wp(a_1)$, $B_1\stackrel{def}{=}\wp_t(a_1)$, $A_2\stackrel{def}{=}\wp(a_2)$, $B_2\stackrel{def}{=}\wp_t(a_2)$. The Laurent series for the function \eqref{HH_SO_General3_4} is the following
\begin{equation}
\begin{gathered}
\label{HH_SO_General_LS3_4}y(t)=\frac{r_i}{t^2}+\frac{b_1}{t}+2r_i(A_1+A_2)+h_0+\left\{(b_1+b_2)A_2-b_2A_1-r_i\left(B_1+B_2\right)\right\}t+\ldots\hfill \\
y(t)=\frac{r_i}{(t-a_1)^2}+\frac{b_2}{t-a_1}+\nu_2+\left\{r_i(B_1+P_{1,2})+(b_1+b_2)T_{1,2}-b_1A_1\right\}(t-a_1)+\ldots \hfill \\
y(t)=\frac{r_i}{(t-a_2)^2}-\frac{b_1+b_2}{t-a_2}+\nu_3+\left\{r_i(B_2-P_{1,2})-b_2T_{1,2}-b_1A_2\right\}(t-a_2)+\ldots, \hfill
\end{gathered}
\end{equation}
where $\nu_2=r_i(2A_1+A_2+T_{1,2})-(b_1+b_2)J_{1,2}+h_0$ and $\nu_3=r_i(A_1+2A_2+T_{1,2})-b_2J_{1,2}+h_0$. In order to find sixth -- order elliptic solutions explicitly we substitute the series \eqref{HH_SO_General_LS1_2} or \eqref{HH_SO_General_LS3_4} into the original equation and set to zero the coefficients at negative and zero powers of $\xi$  in the resulting expression. Note that $\xi=t$ for the first series in \eqref{HH_SO_General_LS1_2} or \eqref{HH_SO_General_LS3_4}, $\xi=t-a_1$ for the second series in \eqref{HH_SO_General_LS1_2} or \eqref{HH_SO_General_LS3_4}, $\xi=t-a_2$ for the third series in \eqref{HH_SO_General_LS1_2} or \eqref{HH_SO_General_LS3_4}. Further one should check that all the Laurent series in a neighborhood of the poles $t=0$, $t=a_1$, $t=a_2$ are distinct. Consequently, in the case of elliptic solutions \eqref{HH_SO_General3_4} we should consider the possibilities given in table \ref{T:HH_SO2}. Indeed, the series are distinct if $b_1\neq 0$ or (and) $b_2\neq 0$, otherwise  supposing that the corresponding coefficients $c_0$, $c_1$ of the series \eqref{HH_SO_General_LS3_4} with $b_1=0$, $b_2=0$ are equal, we see that all other coefficient  at corresponding powers also coincide. Forming an algebraic system we need five coefficients in each of the series $c_k$, $k=0$, $\ldots$, $4$. As a result we obtain $15$ nontrivial algebraic equations. In addition we include into the system the relations $B_i^2=4A_i^3-g_2A_i-g_3$, $i=1$, $2$, $P_{1,2}^2=4T_{1,2}^3-g_2T_{1,2}-g_3$ and expressions given by addition formulae \eqref{Addition_formulae} or \eqref{Addition_formulae_double}. We use the latter expressions  when most of other equations are solved. If we find that $A_1=A_2$  then we take equalities \eqref{Addition_formulae_double}, otherwise~--~\eqref{Addition_formulae}.

\begin{table}[h]%[h]
    \caption{Values of $\sigma$ with the Fuchs indices $j=1$ or (and)~$j=2$ or (and)~$j=3$.} \label{T:HH_SO2}
    \center
       \begin{tabular}[pos]{|c|c|c|c|}
                \hline
                Values of $\sigma$ & Fuchs indices & Type of series & $i$ \\
                \hline
$-\frac43$ & $ -1$, $1$, $4$, $10$ & $y^{(1)}(t)$ & $1$ \\
\hline
$-1$ & $ -1$ ,$2$, $3$, $10$ & $y^{(1)}(t)$ & $1$ \\
\hline
$-\frac{16}{5}$ & $ -1$ ,$1$, $5$, $9$ & $y^{(2)}(t)$ & $2$ \\
\hline
$-16$ & $ -1$, $3$, $5$, $7$ & $y^{(2)}(t)$ & $2$ \\
\hline
$-6$ & $ -1$, $2$, $5$, $8$ & $y^{(2)}(t)$ & $2$ \\
\hline
        \end{tabular}
\end{table}

Some of obtained elliptic solutions are rather cumbersome that is why we present here not complete list of sixth--order elliptic solutions. In the case $\sigma=-1$, $\alpha_2=\alpha_1$ there exist elliptic solutions of the form \eqref{HH_SO_General1}. We set $r_1=-3$, $r_2=-6$ and use addition formulae to obtain the following explicit expression
\begin{equation}
\begin{gathered}
\label{HH_SO_Sol1}y(t)=6\wp(t;g_2,g_3)-\frac{3}{4}\left[\frac{\wp_t(t;g_2,g_3)+B_1}{\wp(t;g_2,g_3)-A_1}\right]^2-\frac{3}{2}\left[\frac{\wp_t(t;g_2,g_3)+B_2}{\wp(t;g_2,g_3)-A_2}\right]^2+h_0,
\end{gathered}
\end{equation}
where the parameter $h_0$, $A_1$, $A_2$, $g_2$, $g_3$, $B_1$, and $B_2$ are given by
\begin{equation}
\begin{gathered}
\label{HH_SO_Sol1_g}h_0=6A_2-\frac{\alpha_1}{2},\quad A_1=0,\quad A_2= \pm\frac{\sqrt{33}\alpha_1}{132},\quad g_2=\frac{\alpha_1^2}{132},\quad g_3=\frac{H}{234}-\frac{\alpha_1^3}{2808},\\
 B_1=\sqrt{\frac{\alpha_1^3}{2808}-\frac{H}{234}},\quad B_2=-B_1
\end{gathered}
\end{equation}
Note that we may take any value of the square root in the expression for $B_1$. Along with this we suppose that $H\neq0$ if $\alpha_1=0$.

Equation \eqref{HH_eqn} possesses sixth--order elliptic solutions of the form \eqref{HH_SO_General2}. With the help of addition formulae we rewrite such a family of elliptic solutions  as
\begin{equation}
\begin{gathered}
\label{HH_SO_Sol2}y(t)=-\frac{3(\sigma+4)}{\sigma}\wp(t;g_2,g_3)+\frac{3\left[\left\{\wp_t(t;g_2,g_3)\right\}^2+B_1^2\right]}{\sigma \left[\wp(t;g_2,g_3)-A_1\right]^2}+h_0.
\end{gathered}
\end{equation}
Note that in this case $A_1=A_2$. The parameter $h_0$, $A_1$, $B_2$, $g_2$, $g_3$ take the form
\begin{equation}
\begin{gathered}
\label{HH_SO_Sol2_g}h_0=\frac{\alpha_1}{4\nu_1\sigma}\left\{ (\sigma^2+24\sigma+32)\kappa-\sigma^{4}-27\sigma^{3}+176\sigma^{2}+384\sigma+512\right\},\\
A_1=-\frac{\alpha_1}{12\nu_1}\left\{\kappa+\sigma^{2}+17\sigma+16\right\}\{\sigma+4\},\quad B_1^2=\frac{\alpha_1^3\sigma\nu_2}{\nu_1^3}\left\{\kappa+\sigma^{2}+17\sigma+16\right\}\\
g_2=\frac{\alpha_1^2\nu_2}{12\nu_1^2} \left\{ \sigma^{2}+32\sigma+16 \right\},\quad
g_3=\frac{\alpha_1^3\nu_2}{216\nu_1^3} \left\{\kappa+\sigma^{2}+17\sigma+16\right\}.
\end{gathered}
\end{equation}
This family of elliptic solutions satisfies equation \eqref{HH_eqn} under two constraints on the parameters of the original equation:
\begin{equation}
\begin{gathered}
\label{HH_SO_Sol2_alpha}\alpha_2=\frac{\alpha_1}{\nu_1}\left\{\left( \sigma+1 \right) \left( \sigma+16 \right)\kappa+ 181\sigma^{2}+272\sigma+256 \right\}, H=\frac{\alpha_1^3}{12\nu_1^3}\left\{76800\nu_3-1891\sigma^7\right.\\
\left.-(181\sigma^{6}+3830\sigma^{5}+77440\sigma^{4}+967960\sigma^{3}+4542080\sigma^{2}+6952576\sigma+3594240)\kappa\right\}.\hfill
\end{gathered}
\end{equation}
In relations \eqref{HH_SO_Sol2_g}, \eqref{HH_SO_Sol2_alpha} the parameters $\kappa$, $\nu_1$, $\nu_2$, $\nu_3$ are given by
\begin{equation}
\begin{gathered}
\label{HH_SO_Sol2_nu}\kappa=\pm \sqrt{181\sigma^{2}+272\sigma+256},\quad \nu_1=\sigma^3+34\sigma^2+140\sigma+272, \\
\nu_2=2(\sigma^{2}+17\sigma+16)\kappa+ \sigma^{4}+34\sigma^{3}+502\sigma^{2}+
816\sigma+512, \\
\nu_3=-\frac {121}{12800}\sigma^{6}+\frac{6187}{640}\sigma^{5}+\frac {314131}{1920}\sigma^{4}+\frac{7115}{8}\sigma^{3}
+\frac {365171}{200}\sigma^{2}+\frac {133556}{75}\sigma+\frac {3744}{5}.
\end{gathered}
\end{equation}
In addition we suppose that $\nu_1\neq0$ and $\sigma\neq-2$. If $\sigma=-2$, then the corresponding coefficients of the Laurent series in a neighborhood of the points $a_1$, $a_2$ coincide and solution \eqref{HH_SO_Sol1} is no longer a sixth--order elliptic function.

Further let us take $\sigma=-6$. We obtain elliptic solutions of the form \eqref{HH_SO_General3_4} with $i=2$, $r_2=-1$. Rewriting this function with the help of addition formulae \eqref{Addition_formulae}, we get
\begin{equation}
\begin{gathered}
\label{HH_SO_Sol3}y(t)=\wp(t;g_2,g_3)-\frac{1}{4}\left[\frac{\wp_t(t;g_2,g_3)+B_1}{\wp(t;g_2,g_3)-A_1}\right]^2-\frac{1}{4}\left[\frac{\wp_t(t;g_2,g_3)+B_2}{\wp(t;g_2,g_3)-A_2}\right]^2+h_0,
\end{gathered}
\end{equation}
where the parameters $h_0$, $B_2$, $g_2$, $g_3$, $B_1$ are given by
\begin{equation}
\begin{gathered}
\label{HH_SO_Sol2_g}h_0=A_1+A_2-\frac{4\alpha_1+\alpha_2}{20},\quad B_2=-B_1,\quad  g_2=\frac {\alpha_2^2}{140}+\frac {4\alpha_1^2}{35}-\frac {4\alpha_2\alpha_1}{105},\\
g_3=\frac {8\alpha_1^3}{675}-\frac {\alpha_1\alpha_2^2}{675}-\frac {4\alpha_1^2\alpha_2}{675}+\frac {\alpha_2^3}{5400}+\frac {2H}{27},\\
B_1^2=4A_2^3+ \left(\frac {4\alpha_1\alpha_2}{105}-\frac {\alpha_2^2}{140}-\frac {4\alpha_1^2}{35}\right)A_2-\frac {8\alpha_1^3}{675}+
\frac{\alpha_1\alpha_2^2}{675}+\frac {4\alpha_1^2\alpha_2}{675}-\frac {\alpha_2^3}{5400}-\frac {2H}{27}
\end{gathered}
\end{equation}
The parameter $A_2$ is arbitrary and the parameter $A_1$ is the following
\begin{equation}
\begin{gathered}
\label{HH_SO_Sol2_A1}A_1=-\frac{A_2}{2}\pm\sqrt{\frac{\alpha_2}{560}+\frac{\alpha_1^2}{35}-\frac{\alpha_1\alpha_2}{105}-\frac{3A_2^2}{4}}.
\end{gathered}
\end{equation}

Again we note that the sixth--order elliptic solutions obtained in this section degenerate if the following condition $g_2^3-27g_3^2=0$ is valid. Family \eqref{HH_SO_Sol2} is given here for the first time, while families \eqref{HH_SO_Sol1}, \eqref{HH_SO_Sol3} are reductions of known general solutions of equation \eqref{HH_eqn} with $\sigma=-1$, $\sigma=-6$.

\section {Conclusion}

In this article we have studied the problem of finding exact
elliptic solutions of a fourth--order ordinary differential equation arising in the H\'{e}non -- Heiles model.
We have given a detailed description of a method enabling one to find all the families of elliptic solutions
of an autonomous ordinary differential equation or a system of autonomous ordinary differential equations.

The fourth--order equation related the H\'{e}non -- Heiles model admits elliptic solutions of even orders only. We have classified  all the families of second--order, fourth--order, and sixth--order elliptic solutions of the equation in question.
We have given explicit expressions for all the families of   second--order and fourth--order elliptic solutions and for dome families of sixth--order elliptic solutions.

The method described in this article generalizes several other
methods, such as the Weierstrass elliptic--function method \cite{Kudr90a, Kudr90b, Kudr91}, the Jacobi elliptic--function method \cite{Parkes02, Fu02, Kudr05} and their
different extensions and modifications \cite{Kudr92, Kudr06}. Most of the methods do not alow one to obtain all elliptic solutions. The only exception is the method of Conte and Musette \cite{Conte02,Conte03}. Since there is no need to find and integrate any additional equation as in the method of Conte and Musette, it seems that the method used in this article is more simple in application.  The main ideas of the method we have described in this article is the following. One uses the local singularity analysis in order to construct an explicit expression for a solution and then  by direct calculations one  finds all the parameters of the solution and constrains on the parameters of the original equation if any. Let us note that Vernov and Timoshkova considered the problem of finding fourth--order elliptic solutions of the equation in question in the cases $\sigma=-4/3$, $\sigma=-16/5$ \cite{Vernov02}. They used the method developed by Conte and Musette and found a first -- order equation satisfied by such families of elliptic solutions, but the did not present these solutions explicitly.

In conclusion we would like to mention that our approach \cite{Demina01, Demina02} may be used to find and to classify simply periodic solutions of the equation in question. This problem will be a topic of further investigations.

\section {Acknowledgements}

This research was partially supported by Federal Target Programm
"Research and Scientific - Pedagogical Personnel of innovation in
Russian Federation on 2009 -- 2013 (Contract P1228).


\begin{thebibliography}{99}


\bibitem{Henon01} \textit{H\'{e}non M., Heiles C.} The applicability of the third integral of motion: some numerical
experiments. Astron. J. -- 1964. --Vol. 69. --P. 73--79.

\bibitem{Weiss01} \textit{Chang Y.F., Tabor M., Weiss J.} Analytic structure of the H\'{e}non -- Heiles Hamiltonian in integrable and
nonintegrable regimes. J. Math. Phys. -- 1982. -- Vol. 23(4). -- P. 531--538.

\bibitem{Weiss02} \textit{Weiss J.} B\"{a}cklund Transformation and Linearizations of the H\'{e}non -- Heiles System.
Phys. Lett. A. -- 1984. -- Vol. 102. -- P. 329--331.

\bibitem{Weiss03} \textit{Weiss J.} B\"{a}cklund Transformation and the H\'{e}non -- Heiles System.
Phys. Lett. A. -- 1984. -- Vol. 105. -- P. 387--389.

\bibitem{Fordy01} \textit{Fordy A.P.} The H\'{e}non -- Heiles System Revisited. Physica D. -- 1991. -- Vol. 52. -- P. 204--210.

\bibitem{Conte01} \textit{Conte R., Musette M., Verhoeven C.} Explicit integration of the H\'{e}non -- Heiles Hamiltonians. J. Non. Math. Phys. -- 2005. -- Vol. 12 (Supplement 1). -- P.  212--227.

\bibitem{Hone02} \textit{Hone A.N.W.} Non--autonomous H\'{e}non -- Heiles systems. Physica D. -- 1998. -- V. 118 (1--2). --
P. 1--16.

\bibitem{Newell01} \textit{Newell A.C., Tabor M., Zeng Y. B.}, A Unified Approach to Painlev´e Expansions. Physica
D. -- 1987. -- Vol. 29. -- P. 1–-68.

\bibitem{Cosgrove01} \textit{Cosgrove C.M.} Higher--order Painlev\'{e} equations in the polynomial class I. Bureau
symbol P2. Stud. Appl. Math. -- 2000. -- Vol. 104. -- P. 1--65.

\bibitem{Cosgrove02} \textit{Cosgrove C.M.} Higher--order Painlev\'{e} equations in the polynomial
class II: Bureau Symbol P1. Stud. Appl. Math. -- 2006. -- Vol. 116. -- P. 321--413.

\bibitem{Eilbeck01} \textit{Eilbeck J.C., Enol'skii V.Z.} Elliptic Baker--Akhiezer functions and an application to an integrable dynamical system. J. Math. Phys. 1994. -- Vol. 35. -- P. 1192--1201 .

\bibitem{Eilbeck02} \textit{Eilbeck J.C., Enol'skii V.Z.} Elliptic solutions and blow--up in an integrable H\'{e}non -- Heiles system. Proceedings of the Royal Society of Edinburgh. -- 1994. -- Vol. 124A. -- P. 1151--1164.


\bibitem{Timoshkova01} \textit{Timoshkova E.I.} A New Class of Trajectories of Motion in the H\'{e}non -- Heiles Potential
Field. Rus. Astron. J. -- 1999. -- Vol. 76. -- P. 470--475.

\bibitem{Vernov01} \textit{Vernov S. Yu.} Construction of solutions for the generalized H\'{e}non -- Heiles system with the help of the Painlev\'{e} test.
Preprint SINP MSU 2002--21/705.

\bibitem{Vernov02} \textit{Vernov S. Yu., Timoshkova E.I.} On two nonintegrable cases of the generalized Henon--Heiles system with an additional nonpolynomial
term. Phys. Atom. Nucl. -- 2005. -- Vol. 68. -- P. 1947--1955.

\bibitem{Eremenko01} \textit{Eremenko A.} Meromorphic traveling
wave solutions of the Kuramoto-Sivashinsky equation. J. Math. Phys.,
Anal., Geom. -- 2006. -- Vol. 2, No. 3. -- P. 278--286.

\bibitem{Demina01} \textit{Demina M.V., Kudryashov N.A.} Explicit expressions for
meromorphic solutions of autonomous nonlinear ordinary differential
equations. Commun. Nonlinear Sci. Numer. Simulat. -- 2011. -- Vol.
16. P. 1127--1134.

\bibitem{Demina02} \textit{Demina M.V., Kudryashov N.A.} From Laurent series to exact
meromorphic solutions: The Kawahara equation. Phys. Lett. A. --
2010. -- Vol. 374. P. 4023--4029.

\bibitem{Demina03} \textit{Demina M.V., Kudryashov N.A.} On elliptic solutions of nonlinear ordinary differential equations. Applied Mathematics and Computation 2011. -- Vol. 217 (23). P. 9849--9853.

\bibitem{Soukharev01} \textit{Kudryashov N.A., Soukharev M.B., Demina M.V.} Elliptic traveling waves of the Olver equation.
Comm. Non. Sci. Num. Simulat. -- 2012. -- Vol. 17(11). -- P. 4104--4114.

\bibitem{Conte02} \textit{Musette M., Conte R.} Analytic solitary waves of nonintegrable equations. Physica D. -- 2003. -- Vol. 181. -- P. 70--79.

\bibitem{Conte03} \textit{Conte R., Musette M.} Elliptic general analytic solutions. Studies in Applied Mathematics. -- 2009. -- Vol.123. -- P. 63--81.




\bibitem{Kudr90a} \textit{Kudryashov N.A.}  Exact solutions of the generalized Kuramoto--Sivashinsky
equation. Phys. Lett. A. -- 1990. -- Vol. 147. -- P. 287--291.

\bibitem{Kudr90b} \textit{Kudryashov N.A.} Exact solutions of the non--linear
wave equations arising in mechanics. J. Appl. Math. Mech. -- 1990.
-- Vol. 54:3. -- P. 372--375.

\bibitem{Kudr91} \textit{Kudryashov N.A.} On types of nonlinear nonintegrable
differential equations with exact solutions. Phys. Lett. A. -- 1991. -- Vol. 155.
-- P. 269--275.

\bibitem{Parkes02}  \textit{Parkes E.J.,  Duffy B.R., Abbott P.C.}
The Jacobi elliptic--function method for finding periodic--wave
solutions to nonlinear evolution equations. Phys. Lett. A. -- 2002.
-- Vol. 295. -- P. 280--286.

\bibitem{Fu02}  \textit{Fu Z.,  Liu S., Liu S.}
New transformations and new approach to find exact solutions to nonlinear equations. Phys. Lett. A. -- 2002. -- Vol. 229. -- P. 507--512.

\bibitem{Kudr05} \textit{Kudryashov N.A.} Simplest equation method to look for exact solutions
of nonlinear differential equations. Chaos, Solitons and Fractals.
-- 2005. -- Vol. 24. -- P. 1217--1231.



\bibitem{Kudr92} \textit{Kudryashov N.A.} Partial differential equations with
solutions having movable first -- order singularities. Phys.
Lett. A. -- 1992. -- Vol. -- 169. -- P. 237--242.

\bibitem{Kudr06} \textit{Kudryashov N.A., Demina M.V.} Polygons of differential equations for finding exact solutions.  Chaos, Solitons and Fractals. -- 2007. -- Vol. 33. -- P.
1480--1496.


%\bibitem{Hone01} \textit{Hone A.N.W.} Non--existence of elliptic travelling wave solutions of the
%complex Ginzburg--Landau equation. Physica D. -- 2005. -- V. 205. --
%P. 292--306.




\end{thebibliography}
\end{document}